\documentclass{appolb}
\usepackage{graphicx}
\usepackage{authblk}

\graphicspath{{./graphics/}}

\newcommand{\eq}[1]{\begin{equation} #1 \end{equation}}

\begin{document}
\title{
\uppercase{Cumulative pion production via successive collisions in nuclear medium}
\thanks{Presented at ``Critical Point and Onset of Deconfinement'', Wroc\l{}aw, Poland, May 30th -- June 4th, 2016}%
}
\author[1]{A.~Motornenko}
\author[2,3]{M. I. Gorenstein}
\affil[1]{Taras Shevchenko National University of Kyiv, Kyiv, Ukraine}
\affil[2]{Bogolyubov Institute for Theoretical Physics, Kyiv, Ukraine}
\affil[3]{Frankfurt Institute for Advanced Studies, Johann Wolfgang Goethe University, Frankfurt, Germany}

\maketitle
\begin{abstract}
Production of pions in proton-nucleus (p+A) reactions outside of a kinematical boundary of proton-nucleon collisions,
the so-called cumulative effect, is studied. The kinematical restrictions on pions emitted in backward direction
in the target rest frame are analyzed. It is shown that cumulative pion production requires
a presence of massive baryonic resonances that are produced during successive collisions of projectile with nuclear nucleons.
After each successive collision the mass of created resonance may increase and, simultaneously, its
longitudinal velocity decreases. Simulations within Ultra relativistic Quantum Molecular
Dynamics model reveals that successive collisions of baryonic resonances with nuclear nucleons plays the
dominant role in cumulative pion production in p+A reactions.

\end{abstract}
\PACS{25.75.-q, 25.75.Dw, 25.75.Ld}

\section{Introduction}
Cumulative effect in proton-nucleus (p+A) reactions is a production of secondary particles in a kinematic region
forbidden in proton-nucleon (p+N) collisions at the same energy of projectile protons. First experiments with detection of
cumulative particles were performed at Synchrophasotron accelerator of the Joint Institute for Nuclear Research in Dubna \cite{baldin-74, baldin-77, leksin-77}.
In this work inclusive reactions p+A$\rightarrow \pi(180^\circ)+X$ are considered with pions emitted in
backward direction, i.e. at 180$^\circ$, in the target rest frame.

Let $E_{\pi}^{*}$ denotes the maximal possible energy of the pion emitted
at angle $180^{\circ}$ in the laboratory frame
in p+N interaction at fixed projectile proton
momentum $p_0$. In p+A collisions at the same projectile proton momentum $p_0$,
pions emitted at $180^{\circ}$
in the
the nucleus rest frame with energies $E_{\pi}> E_{\pi}^*$,
even above 2$E_{\pi}^*$, were experimentally observed \cite{baldin-74,baldin-77,leksin-77, frankel-79, bayukov-79}.

The main physical quantities in our analysis are the masses and longitudinal
(i.e. along the collision axis) velocities
of the baryonic resonances created in p+A reactions.
Resonance $R$ is first produced in p+N$\rightarrow R$+N reaction,
and it then participates in successive $R$+N$\rightarrow R'$+N collisions.
Due to subsequent collisions of the resonances
with nuclear nucleons  the resonance mass may increase and its longitudinal
velocity decreases.
We argue that the cumulative pions in p+A reactions are created
by baryonic resonances with very high masses that are formed due to successive collisions
with nuclear nucleons.
We also use the UrQMD model to
analyze some microscopic aspects of cumulative pion production in p+A reactions.

\section{Successive collisions with nuclear nucleons}\label{sec-multi}
The different theoretical models were proposed
to describe the cumulative pion production.
However, origin of this effect is still not settled.
In the present study we will advocate the approach suggested in Ref.~\cite{gor-77}.
More details are presented in our recent paper \cite{MG}, where the references
to other theoretical models can be also found.
We assume that cumulative
particle production takes place due to the large mass of the projectile
baryonic resonance created in the {\it first} p+N collision
and its further propagation through the nucleus.
This baryonic resonance has a chance to interact with other nuclear nucleons
earlier than it decays to free final hadrons.
It can be shown \cite{MG} that during
{\it successive collisions} of the baryonic resonance with nuclear nucleons
it is possible both to enlarge the resonance mass
$M_R$ and, simultaneously, to reduce its longitudinal velocity $v_R$.

Production of any additional hadron(s) and/or a presence of non-zero
transverse momenta in the final state would require an additional energy
and lead to a reduction of $E^{\rm max}_\pi$ value at fixed projectile momentum $p_0$.
Thus, to find the maximum of pion energy one should
restrict the kinematical analysis to the one-dimensional
(longitudinal) direction, i.e., all particle momenta should be directed along the collision axis.

If one consider baryonic resonance decay, $R\rightarrow {\rm N}+\pi(180^\circ)$,
the value of $E_\pi$ depends on the resonance mass $M_R$ and its longitudinal velocity $v_R$.
In the resonance rest frame the pion energy and momentum can be easily found
\eq{\label{E0-R}
E^{0}_\pi= ~\frac{M_R^2-m_N^2+m_\pi^2}{2M_R}~,~~~~~p_\pi^{0}=\sqrt{(E_\pi^{0})^2-m_\pi^2}~.
}
The energy $E_\pi$ (with neglected pion mass for simplicity), in the laboratory frame, is obtained as
\eq{\label{Epi-lab}
E_\pi= ~\frac{E^{0}_\pi - v_R p_\pi^{0}}{\sqrt{1-v_R^2}}~.
}
Therefore, both the increase of $M_R$
and decrease of $v_R$ provide an extension of the available
kinematic region of $E_\pi$ for pions emitted at $180^{\circ}$.
Thus, the suppression of $E_\pi$ compared to $E^{0}_\pi$ can be interpreted as Doppler
(``red shift'') effect.

As seen from Eq.~(\ref{Epi-lab}), both effects of resonance mass increase and its velocity decrease lead to larger values of $E_\pi$ and, thus,
extend
the kinematic region for cumulative pion production.
The object responsible for the cumulative production of $\pi(180^\circ)$, i.e.,
the heavy and slow moving resonance, does not exist
inside a nucleus but is formed during the whole evolution process of p+A reaction.

Let us consider successive collisions with nuclear nucleons:
${\rm p}+{\rm N}\rightarrow R_1+ {\rm N}$, $R_1+{\rm N}\rightarrow R_2+ {\rm N}$,~...~, $R_{n}+{\rm N}\rightarrow {\rm N}+{\rm N}+\pi(180^\circ)$.
The nuclear nucleons are considered as free particles. This approximation
can be justified by the fact that the projectile proton energy is typically
3 orders of magnitude larger than the binding energy of nucleons in a nucleus.
It is assumed that after $n$-th collision the baryonic resonance decays, $R_n\rightarrow \pi(180^\circ)$ + N.
The energy and momentum conservation between initial and final state read as
\eq{\label{cons-n}
\sqrt{m_N^2+p_0^2}+n\cdot m_N =\sum_{i=1}^{n+1}\sqrt{m_N^2+p_i^2}+E_\pi~,~~~~~~
p_{0}=\sum_{i=1}^{n+1}p_i-p_\pi~.
}
The maximal pion energy $E_\pi$ after $n$ successive collisions
denoted now $E_{\pi,n}^*$ can be found
from Eq.~(\ref{cons-n}) using the extremum conditions
$\partial E_\pi /\partial p_{i}=0$. This leads to
\eq{\label{p*n}
p_{N,n}^*\equiv p_1=p_2=...=p_{n+1}=\frac{p_0+p^*_{\pi,n}}{n+1}~,
}
and gives an implicit equation for $E_{\pi,n}^*$.

\begin{figure}[h!]
\centering
  \includegraphics[width=0.49\textwidth]{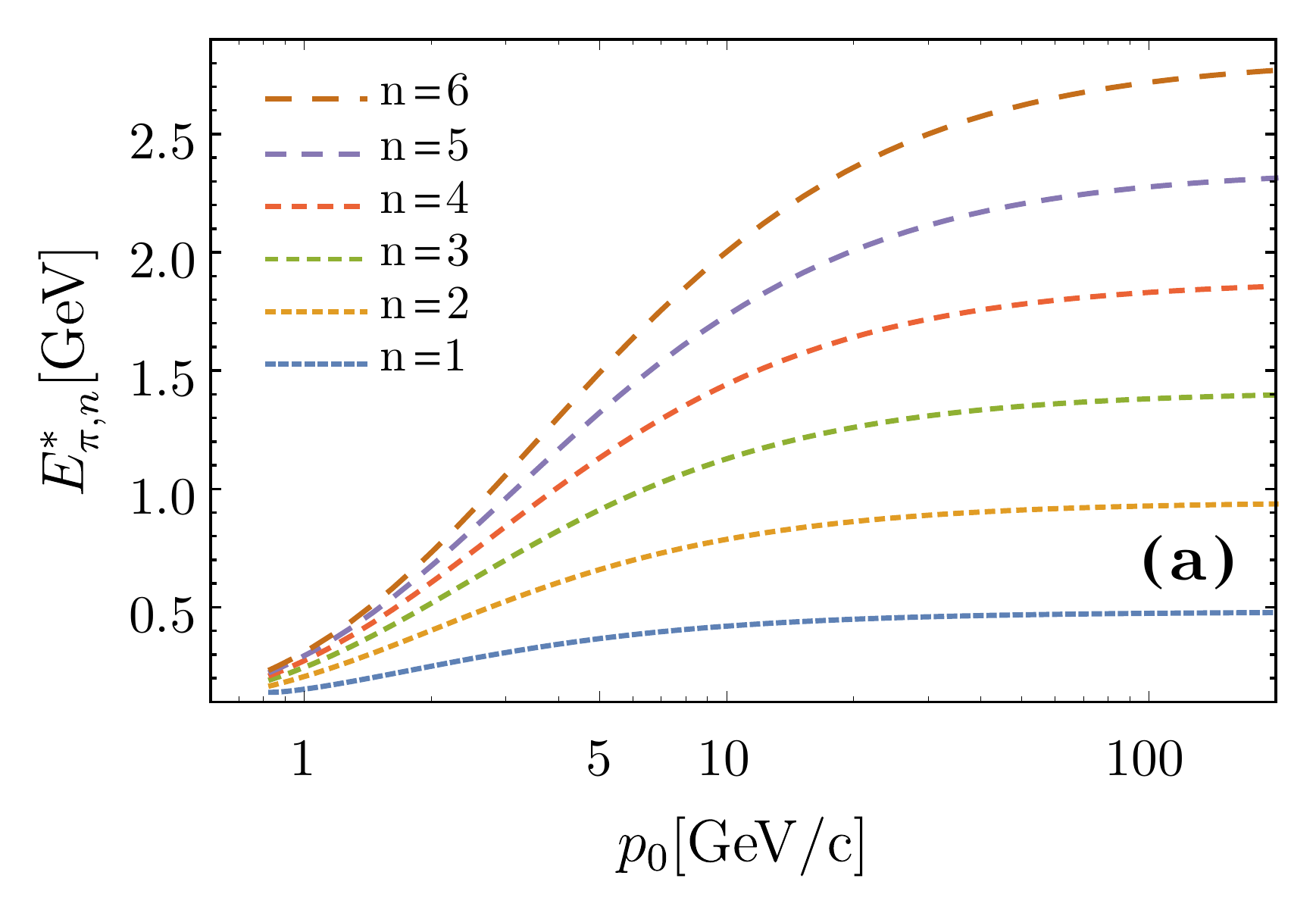}
    \includegraphics[width=0.49\textwidth]{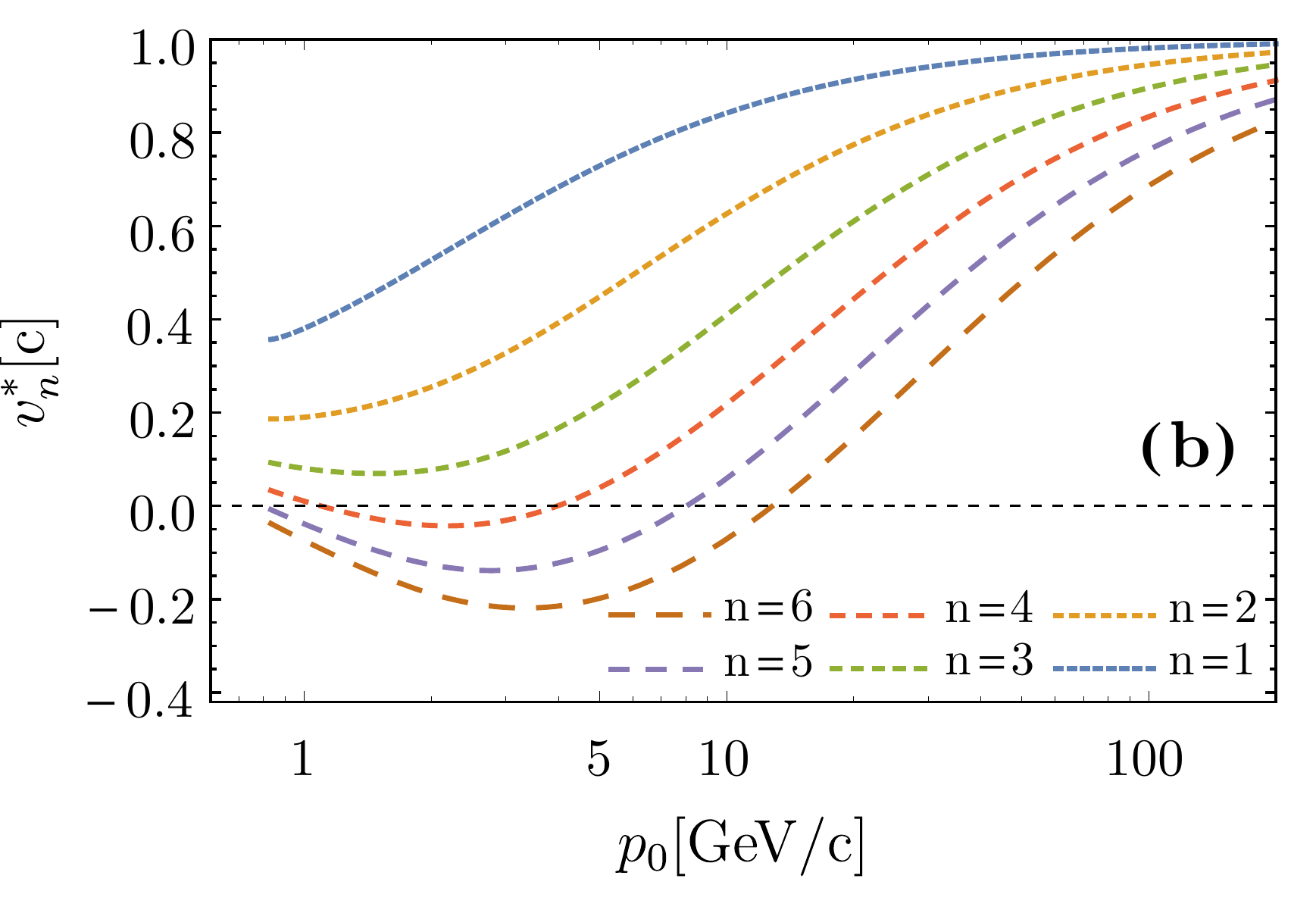}
  \caption{\footnotesize Maximal energies $E_{\pi,n}^*$ (a) and velocities $v_n^*$ (b)
  of the baryonic resonances after $n$ successive collisions with nuclear nucleons.
  The values of $v_n^*$ are required to provide the maximal energy
  $E_{\pi ,n}^*$ of $\pi(180^\circ)$.
  They are calculated from Eq. \ref{cons-n} with assumption of $\partial E_\pi /\partial p_{i}=0$, as functions of projectile proton momentum $p_0$.}
  \label{fig-En}
\end{figure}

The maximal energies $E_{\pi,n}^*$ of pions emitted at $180^{\circ}$ are
presented in Fig.~\ref{fig-En}.
A surprising behavior with $v_n^*<0$ for $n\ge 4$ is observed
at some finite regions of projectile momentum $p_0$, i.e., heavy resonance may start to move backward after a large number of successive collisions for not too large $p_0$.
In p+N$\rightarrow R$+N reactions the only $v_R$ values with $v_R>0$ are permitted.

It should be noted that the values of $E^*_n$ and $v^*_n$ found in this section are by no means typical (or average) ones. In fact, the probability to reach these values in p+A reaction is very small. In other words, cumulative pion production is a very rare process.
\section{UrQMD simulations}\label{sec-UrQMD}
In this section the analysis of the cumulative production of $\pi (180^\circ)$ within the UrQMD model~\cite{urqmd} is presented. The UrQMD gives a unique opportunity to study a history of each individual reaction.
\begin{figure}[h!]
\centering
\includegraphics[width=0.48\textwidth]{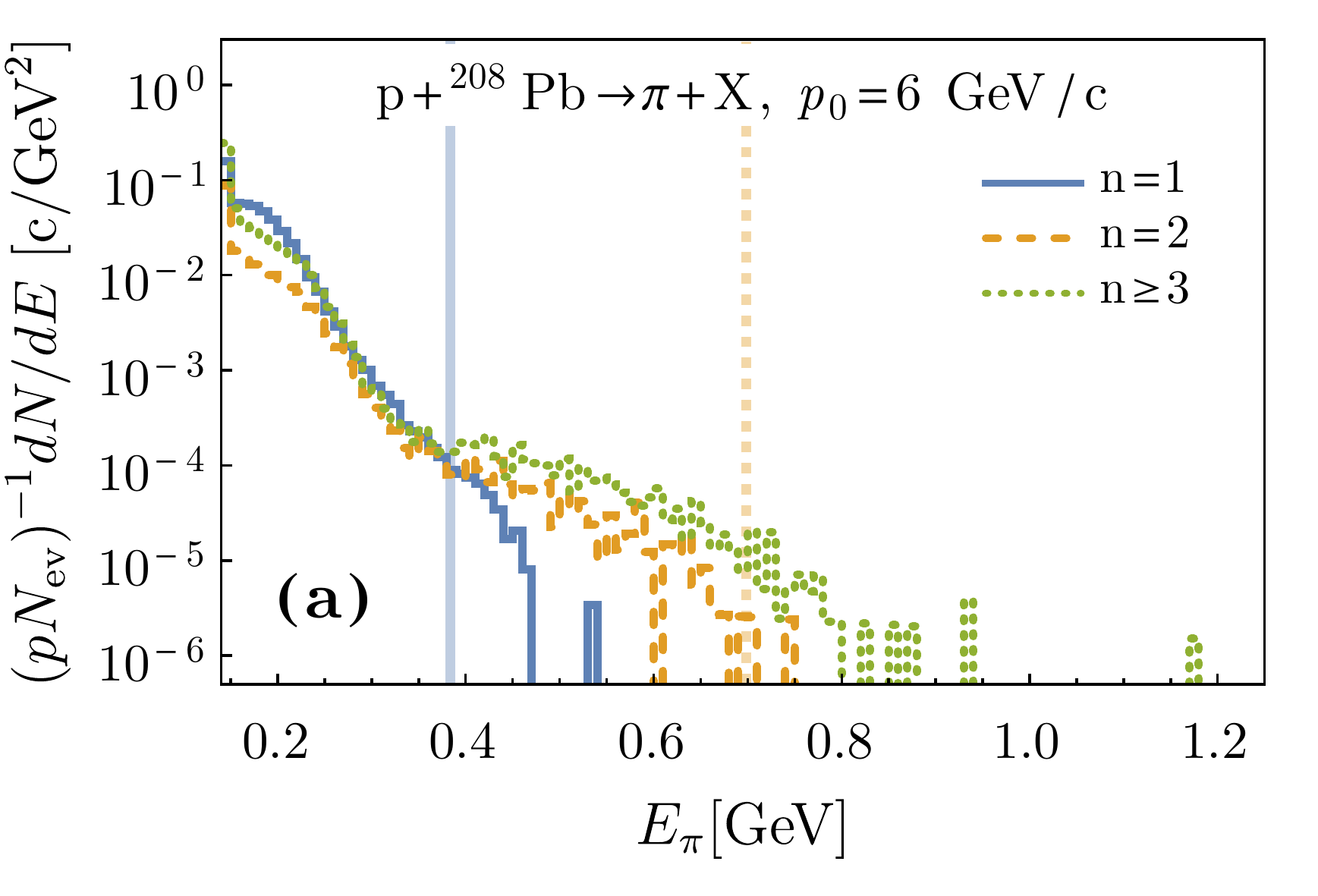}
\includegraphics[width=0.48\textwidth]{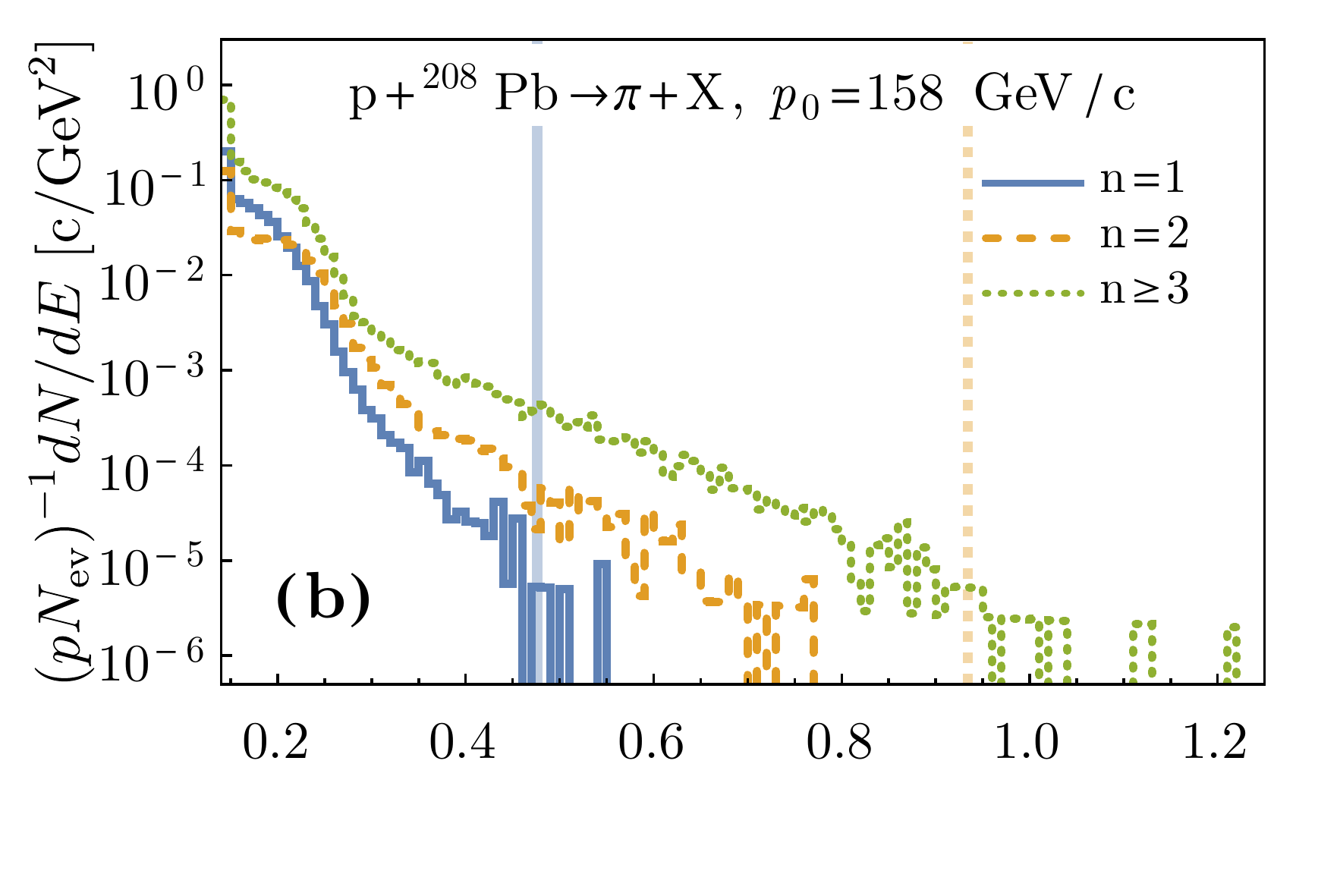}
\caption{\footnotesize{UrQMD results for the pion energy spectra at $180^{\circ}$ in p+$^{208}$Pb collisions. Spectra for pions created after different number of collisions with nuclear nucleons:
$n=1$ (solid line) collision, $n=2$ (dashed line), and $n\ge 3$ (dotted line).
Vertical lines correspond to energies $E^*_{\pi,1}$ (solid) and $E^*_{\pi,2}$ (dashed).
In (a) $p_0=6$~GeV/c and the number of events $N_{\rm ev}=5.7\cdot 10^7$.
In (b) $p_0=158$~GeV/c, $N_{\rm ev}=4.2\cdot 10^7$.}}
\label{fig-pPb}
\end{figure}
In Fig.~\ref{fig-pPb} we compare the spectra of $\pi(180^{\circ})$ emitted from resonance
decay after $n=1$ (solid line), $n=2$ (dashed line),
and $n\geq 3$ (dotted line) successive collisions of the projectile with nuclear nucleons.

From Fig.~\ref{fig-pPb} one observes that $E_\pi$ may exceed $E^*_{\pi,1}$ even
for $n=1$ contribution. This happens because of nucleon motion inside nuclei (Fermi motion)
which exists in the UrQMD model. This effect is, however, not large. The main contribution
to the kinematical region forbidden for p+N collisions (i.e., to $E_\pi > E^*_{\pi,1}$)
comes from the decays of resonances created within $n=2$ and $n\ge 3$ successive collisions with nuclear nucleons.
Therefore, the proposed mechanism of the cumulative pion production -- the successive interactions of heavy resonances with nuclear nucleons --
 is supported by the UrQMD analysis.
\section{Summary}\label{sec-sum}
Pions emitted in p+A reactions at 180$^\circ$ in the target rest frame are considered.
Extension of a kinematical boundary of p+N reactions due to existence of massive baryonic resonances is studied.
These resonances are produced after several successive collisions of projectile with nuclear nucleons:
resonances $R$ created in p+N reactions may have further
inelastic collisions in the nuclear medium. Due to successive
collisions with nuclear nucleons the masses of these resonances may increase
and simultaneously their longitudinal velocities decrease.
These two effects give an explanation of the cumulative pion production.
The simulations of p+A reactions within the UrQMD model support this
physical picture.

\vspace{0.3cm}
{\bf Acknowledgements.} This work is supported by the Goal-Oriented Program
of the National Academy of Sciences of Ukraine and the European Organization of Nuclear
Research (CERN), Grant CO-1-3-2016.

\end{document}